\documentclass[twocolumn,showpacs,amsmath,amssymb]{revtex4}

\usepackage{epsfig}
           
\usepackage{dcolumn}

\usepackage{bm}

\usepackage{amsmath}

\usepackage{amsfonts}

\usepackage{amssymb}

\usepackage{graphicx}%

\newcommand{\beq}{\begin{eqnarray}}

\newcommand{\eeq}{\end{eqnarray}}

\begin{document}

\title{A Green-function approach to transport phenomena in quantum pumps.}

\author{Liliana Arrachea$^{1}$}

\affiliation{Instituto de Biocomputaci\'on y F\'{\i}sica de Sistemas
  Complejos, Universidad de Zaragoza\\
Corona de Arag\'on 42, (50009) Zaragoza, Spain. }


\begin{abstract}
We present a general treatment based on non-equilibrium
Green functions
 to study transport phenomena in systems
described by tight-binding Hamiltonians coupled to reservoirs and 
with one or more time-periodic potentials. We apply this treatment to
the study of transport phenomena in a double barrier structure 
with one and two harmonically time-dependent potentials. Among other properties, 
we discuss the origin of the sign of the net current.

\end{abstract}

\maketitle

\section{Introduction}

The impresive development in the technology of fabrication of small circuits,
enabled the investigation of single electron transport induced by time-periodic
fields \cite{urb}. 
In a recent experiment by Switkes et al \cite{swi}
 charge transport through a quantum dot is induced by means of 
two time-periodic potentials with a phase lag. 
This has renewed the interest in the
study of charge pumping in open quantum systems, motivating 
theoretical \cite{brou,alt1,alt2,ent,moskad1,moskad2,mosk} and experimental  activity  \cite{exp}. 
Closely related phenomena are the photovoltaic effect \cite{photo1,photo2,photo3} in
mesoscopic junctions 
and charge driving in molecular ratchets  \cite{rat,leh}. The main
feature characterizing these effects is the generation of a net current as a
response to a time dependent external field without a net static bias.

Several theoretical treatments on quantum pumps rely on adiabatic approximations 
\cite{brou,alt1,alt2,ent}, relevant for very slow potential modulations.  
A good amount of work on time-dependent transport 
is based on Floquet theory \cite{rat,leh,moskad1,moskad2,mosk,cam,rev}, the 
matrix approach  \cite{brou,ent,ped}
or on the transfer matrix technique \cite{sols}. 
An alternative framework to investigate
transport phenomena in mesoscopic devices and nanostructures is the
description of the device in terms of tight-binding Hamiltonians and
the solution of the problem with non-equilibrium Green functions.
 Since the proposals of Refs.  \cite{caro,meir}, 
this kind of approach became widely extended in the study
of electronic transport through a nanometric or mesoscopic sample, as a response to a
 static bias.  One of the reasons for the success of this scheme is the fact
 that it is suited to 
deal with arbitrary high bias, finite temperature, arbitrary strength of dissipation 
and that it can be extended to include many-body interactions at least pertubatively. 
Another appealing feature of this approach is the possibility of combining it with
the so-called ``ab-initio'' methods to describe  details
of the contacts and molecular bridges of the devices \cite{datta}.
In the context of transport problems with time-dependent fields, there are
also basic proposals of this type of strategies \cite{win,past} but, in
comparison, not so
many recent developments.
Some examples are studies on  ac-driven quantum dots and
superlattices
\cite{plat1, plat2,plat3,plat4}, the latter restricted to weakly coupled
quantum wells, studies on the dynamical Franz-Keldysh effect \cite{jau},
superconducting point contacts with a time-dependent voltage 
\cite{alf}, and conducting rings threaded by a time-dependent  magnetic flux
\cite{lili1,lili2,lili3,lili4}. In these problems, the time-dependent part of the
Hamiltonian is restricted to a single point \cite{plat1, plat2}, bond  
\cite{lili1,lili2,lili3,lili4} or contacts \cite{win,alf}, while approximations
are introduced to deal with more gereral situations \cite{past,plat3,plat4,jau}.

In this work, we present a general treatment based on non-equilibrium
Green functions to study  transport
phenomena in systems described by tight-binding models
in contact with particle reservoirs 
with several time-periodic local potentials. We derive exact equations of
motions and
present  results on the transport 
properties in the special simple cases of one-dimensional systems with one and two 
time-dependent potentials. 
In the first case, some analytical expressions are available. In the
second one, which is relevant for the experimental configuration of
Ref. \cite{swi},  we compute the Green functions numerically.

The traditional and intuitive  way
to think about stationary transport through a mesoscopic device placed between two
electrodes at different chemical potentials is in terms of the behavior of the
density of
states of the central system and of its environment. 
In a time-dependent problem the density of states 
depends on time and its convolution with a Fermi function does not directly
correspond to the notion of occupied energy states. 
In our study  we analyze 
the connection between the transport behavior of the pumps, the density of
states of the environment and the 
 non-equilibrium spectral densities at the positions where the time-dependent potentials are
applied. 

In stationary transport like that resulting as the response to a
static bias, the carriers
responsible for the transport process are those injected from the electrodes 
with energies between the two different chemical potentials. 
A remarkable property of the pumping mechanism is that not only electrons 
with energies close to the Fermi energy of the reservoirs 
contribute to the net electronic current. Instead,  {\em all} the electrons 
contribute to the net flow. This
point has been previously addressed in Ref. \cite{sols} for the problem of a
time-dependent harmonic 
potential in an asymmetric structure. We present here further details
on this behavior which combines effects like photon-assisted tunneling, 
quantum interference and dissipation, sometimes
giving rise to patterns that resemble a turbulent motion of
electrons through the device. 

The control of the direction of the
net current is central for eventual technological applications of the pumping
effect. 
However, the complex nature of the  electronic motion generated in a
quantum pump makes the prediction 
of this property from {\em a priori}  considerations a very
 difficult task. We discuss  some operational conditions where
the sign of the induced current can be understood and we also identify some mechanisms
causing sign reversals.   
The paper is organized as follows. The theoretical treatment is presented
in section II. Examples and results are presented in section III and
IV. Finally, section V is devoted to summary and conclusions.

\section{Theoretical treatment}

\subsection{General model and Green functions}
\begin{figure}
\includegraphics[width=8cm,clip]{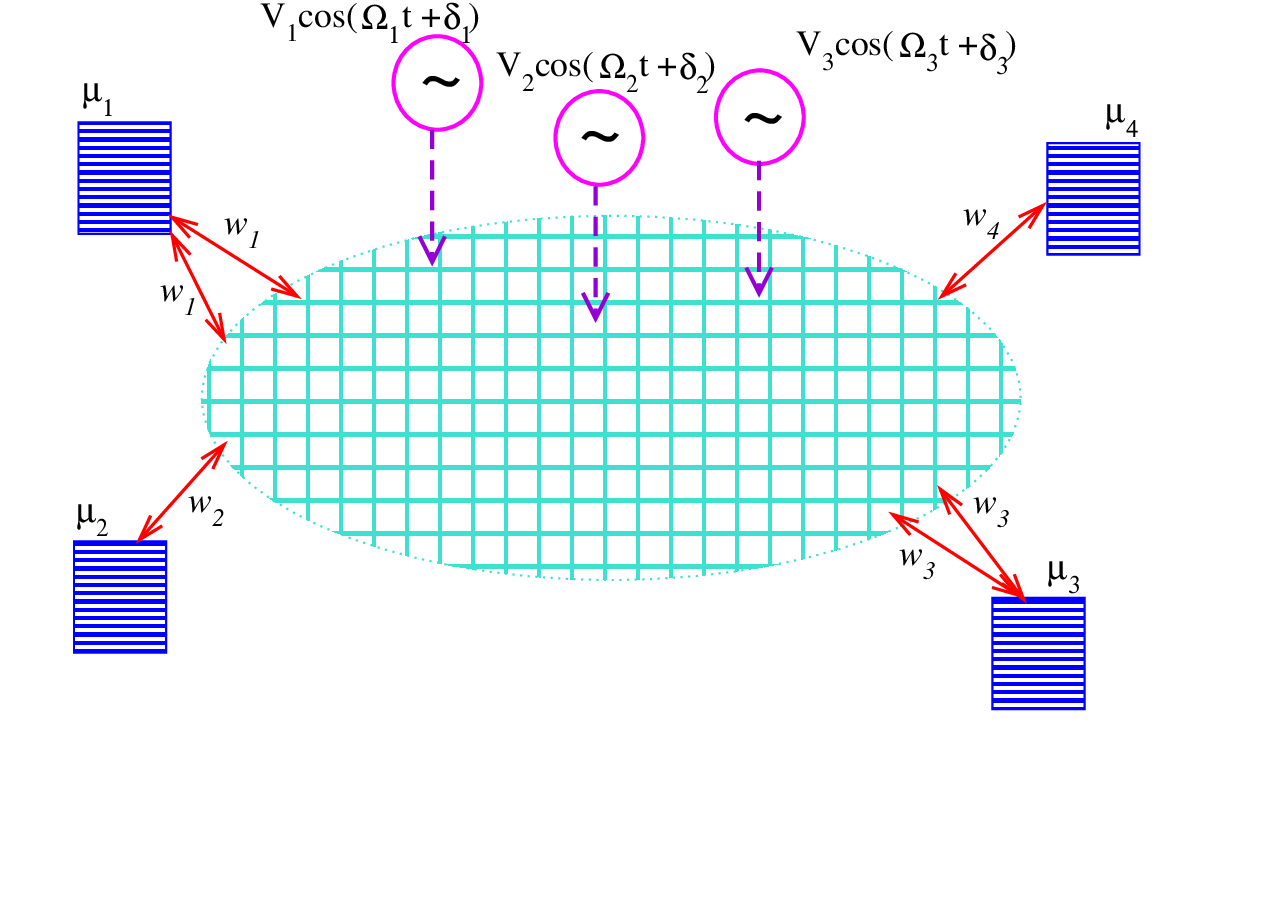}
\caption{(Color online) Scheme of the system described by Hamiltonian
(\ref{genham}) with $L=4$ reservoirs and $M=3$ pumping potentials. 
The grid represents the sites of the central system and the
boxes with stripes represent the reservoirs with chemical potentials
$\mu_{\alpha}$, with $\alpha=1,\ldots,4$.}
\label{fig0}
\end{figure}

We assume a general device, which consists in $L$ reservoirs and a 
 central system driven by several time-periodic local potentials. 
We do not include the effect of many-body interactions in our
treatment. 
The full system is
 described  by the Hamiltonian

\begin{eqnarray}
& & H  =  \sum_{\alpha=1}^{L} H_{\alpha} - \sum_{\alpha=1}^{L} w_{\alpha}
\sum_{k_{\alpha},l_{\alpha}}  
a^{\dagger}_{k_{\alpha}} c_{l_{\alpha}} -
\sum_{\langle lj \rangle } w_{lj} c^{\dagger}_l c_j  + \nonumber \\
& & \sum_{l=1}^N \varepsilon_l  c^{\dagger}_l c_l 
 + \sum_{l=1}^M V_l \cos(\Omega_l t + \delta_l) c^{\dagger}_{i_l} c_{i_l} + H.c.,
\label{genham}
\end{eqnarray}
where the fermionic operators $a_{k_{\alpha}}$ denote degrees of freedom corresponding
to the reservoir $\alpha$, which is described by  the Hamiltonian 
$H_{\alpha}$. The contact of the reservoir $\alpha$ with the 
central region is represented by the hopping element $w_{\alpha}$ between the
reservoir and the 
sites $l_{\alpha}$ placed at the boundaries of the lattice of the central
system. The model for the latter piece is a tight-binding Hamiltonian with $N$ 
lattice positions, hopping
elements $w_{lj}$ between pairs of sites $ \langle lj \rangle$, 
and local energies $\varepsilon_l$. We also consider time-dependent
potentials acting locally at $M$ sites of the central
system denoted by $i_l$, oscillating with amplitudes $V_l$, frequencies $\Omega_l=n
\Omega_0$ (being $n$ an integer number)
and phases $\delta_l$. A scheme of the setup is shown in Fig. \ref{fig0}.

In what follows, we present a closed set of equations to calculate the
Green functions corresponding to spacial coordinates of the central system.
As usual \cite{ram}, we define retarded and lesser Green functions
\begin{eqnarray}
G^R_{l,m}(t,t')&=& -i \Theta(t-t') \langle \{c_l(t), c^{\dagger}_m(t')\}
\rangle, 
\nonumber \\
G^<_{l,m}(t,t')&=& i \langle c^{\dagger}_m(t') c_l(t) \rangle.
\label{green}
\end{eqnarray}
In the non-equilibrium formalism, Dyson equation for the Green function has
a matricial structure which results in a coupled set of integro-differential
equations for the lesser and retarded components. 
Following the standard procedure \cite{caro,meir,datta,win,past} we  consider ideal
non-interacting leads for the reservoirs. The corresponding degrees of freedom are
integrated out, defining the self-energies 
\begin{equation}
\Sigma_{l,m}^{R, <} (t-t')=\delta_{l,l_{\alpha}} \delta_{m,l'_{\alpha}}   |w_{\alpha}|^2 
\sum_{k_{\alpha}} g^{R, <}_{k_{\alpha},k_{\alpha}} (t-t'),
\end{equation}
which depend on the equilibrium Green functions $g^{R, <}_{k_{\alpha},k_{\alpha}} (t-t')$ of the
free reservoirs 
(i.e. isolated
from the central system). The latter 
can be expressed in terms of densities of
states $\rho_{\alpha}(\omega)$ through
\begin{eqnarray}
\sum_{k_{\alpha}}
g^R_{k_{\alpha},k_{\alpha}}(t - t')&=&- i \Theta(t-t')\int_{-\infty}^{\infty} 
\frac{d\omega}{2\pi} e^{-i \omega (t-t')} \rho_{\alpha}(\omega)\nonumber \\
\sum_{k_{\alpha}}
g^<_{k_{\alpha},k_{\alpha}}(t-t')&=&i \int_{-\infty}^{\infty} 
\frac{d\omega}{2\pi} e^{-i \omega (t-t')}
f_{\alpha}(\omega)
 \rho_{\alpha}(\omega),
\end{eqnarray}
 being
$f_{\alpha}(\omega)=1/(e^{\beta(\omega-\mu_{\alpha})}+1)$  the Fermi function.

We work in units where $\hbar=1$.
The equations for the retarded and lesser components read
\begin{eqnarray}
& & \{-i \frac{\partial}{\partial t'} - \varepsilon_n - V_n(t')\} G^R_{m,n}(t,t') 
 - \sum_{l=1}^N G^R_{m,l}(t,t') w_{l,n} \nonumber \\
& & -\sum_{l=1}^{N } \int_{t'}^{t} dt_1  G^R_{m,l}(t,t_1) \Sigma^R_{l,n}(t_1-t') = \delta(t-t') \delta_{m,n}, \nonumber \\
& &\{-i \frac{\partial}{\partial t'} - \varepsilon_n - V_n(t')\} G^<_{m,n}(t,t') 
- \sum_{l=1}^N G^<_{m,l}(t,t') w_{l,n} \nonumber \\
& &-\sum_{l=1}^{M }   [ \int_{-\infty}^{t} dt_1 G^R_{m,l}(t,t_1) \Sigma^<_{l,n}(t_1-t') \nonumber \\
& & + \int_{-\infty}^{t'} dt_1 G^<_{m,l}(t,t_1) \Sigma^A_{l,n}(t_1-t') ]
=0,
\label{dyder}
\end{eqnarray}
being $V_n(t)= \delta_{n,i_l} V_l \cos(\Omega_l t + \delta_l)$,  $l=1,\ldots, M$.

We now present a very convenient way to calculate the Green functions. We
start from the following integrated form of eqs. (\ref{dyder}):
\begin{eqnarray}
 G^<_{m,n}(t,t') & = &  \sum_{i,j=1}^N \int_{-\infty}^t dt_1 
\int_{-\infty}^{t'} dt_2 G^R_{m,i}(t,t_1)  \nonumber \\
& &\times \Sigma^<_{i,j}(t_1-t_2)  G^A_{j,n}(t_2,t'), \nonumber \\
G^R_{m,n}(t,t') & = &  G^0_{m,n}(t-t') + \sum_{i=1}^N \int_{t'}^t dt_1
G^R_{m,i}(t,t_1)  \nonumber \\
& & \times  V_i(t_1) G^0_{i,n}(t_1-t').
\label{dyson}
\end{eqnarray}
 The advanced Green function
is related to the retarded one through $G^A_{j,n}(t,t')=[G^R_{n,j}(t',t)]^*$.
The  equation for the lesser Green function is valid for long enough $t,t'$,
such that no memory 
on the initial condition is
preserved. The ``unperturbed'' retarded Green function $G^0_{m,n}(t-t')$
corresponds to the solution of the problem of the central system {\em without}
time-dependent potentials {\em coupled} to the reservoirs.

We now define the Fourier transform
\begin{equation}
G^R_{m,n}(t,\omega)=\int_{-\infty}^t dt'
G^R_{m,n}(t,t') e^{i (\omega + i \eta)  (t-t')},
\label{four}
\end{equation}
with $\eta=0^+$,
which leads to a linear set for the retarded Green function.
\begin{eqnarray}
& & G^R_{m,n}(t,\omega) = 
G^0_{m,n}(\omega)+ \sum_l \frac{V_l}{2} e^{i (\delta_l+\Omega_l t)}
 \nonumber \\
& & \times G^R_{m,i_l}(t,\omega-\Omega_l) G^0_{i_l,n}(\omega)  + \sum_l \frac{V_l}{2} e^{-i (\delta_l+\Omega_l t)} 
\nonumber \\
& & \times G^R_{m,i_l}(t,\omega+\Omega_l) 
G^0_{i_l,n}(\omega).
\label{dyret}
\end{eqnarray}

 In most of the 
cases, this set must be solved numerically, by discretizing $\omega$ 
 introducing high and low frequency cut-offs 
$ \pm K \Omega_0 $,
respectively. The parameter $K$ depends on $\Omega_0$ and
must satisfy that $|K \Omega_0| $ is much larger than the absolute value of
the highest frequency for which $G^0_{m,n}(\omega)$ has a finite spectral weight.
In the numerical procedure, it must be checked that the solution of
$G^R_{m,n}(t,\omega)$ does not
depend on $K$.

Since $G^R_{m,n}(t,\omega)$ is a periodic function of $t$ with period
$\tau_0=2\pi/\Omega_0$, it is sometimes useful to work with the expansion:
\begin{equation}
G^R_{m,n}(t,\omega)=\sum_{k=-\infty}^{\infty} 
{\cal G}_{m,n}(k,\omega) e^{i k \Omega_0 t},
\label{four2}
\end{equation}
being
\begin{equation}
{\cal G}_{m,n}(k,\omega) = \frac{1}{\tau_0} \int_0^{\tau_0} dt  e^{-i k \Omega_0 t}
G^R_{m,n}(t,\omega).
\end{equation}

We stress that the advantage of writing the Dyson 
equation in the form (\ref{dyson}) and working with the Fourier
transform (\ref{four}) is that the {\em exact} retarded Green function 
can be evaluated from a set of linear equations irrespectively the
amount of time-periodic potentials. In addition, the present formulation in terms of
eq. (\ref{dyret}) is very convenient to perform systematic expansions in powers 
of $V_l$.

\subsection{Pumped current in a one-dimensional system
coupled  to two reservoirs}
We now consider the case where the pumped system  is a tight-binding chain
with hopping elements between nearest neighbors, which is
placed between left and right
reservoirs. The Hamiltonian (\ref{genham}) reduces to
\begin{eqnarray}
H  & = &  H_L + H_R  + H_C(t) -w_L (a^\dagger_L c_1 + H.c.) \nonumber \\
& & -w_R (a^\dagger_R c_N + H.c. ) ,
\label{ham1d}
\end{eqnarray}
with $H_C(t)$  denoting the Hamiltonian for the central piece.

The current from the reservoirs to the central region
can be written as
\begin{equation}
J_{\alpha}(t) = 2 e w_{\alpha} \mbox{Re}[G^<_{i,\alpha}(t,t)],
\end{equation}
with $\alpha=L$ and $i=1$ ($\alpha=R$ and $i=N$)
for the current flowing from the left (right) reservoir.
The dc components of the above currents  are
\begin{equation}
J^{dc}_{\alpha}=\frac{1}{\tau_0}\int_0^{\tau_0} dt J_{\alpha}(t),
\label{cont}
\end{equation}
and due to the continuity condition, they satisfy $J^{dc}_L=-J^{dc}_R$
which allows us the write the dc current $J$ flowing through the central
device as $ J= (J^{dc}_L - J^{dc}_R)/2$.

Following \cite{win} we write
\begin{eqnarray}
J & = &
 \frac{1}{\tau_0} \int_0^{\tau_0} dt
\int_{-\infty}^t dt_1 
\mbox{Re} \{|w_{L}|^2[G^R_{1,1}(t,t_1) g^<_{L}(t_1-t)  + \nonumber\\
& & G^<_{1,1}(t,t_1) g^A_{L}(t_1-t)] - 
 |w_{R}|^2[G^R_{N,N}(t,t_1) g^<_{R}(t_1-t)  \nonumber\\
& & + G^<_{N,N}(t,t_1) g^A_{R}(t_1-t)],
 \},
\label{jdc2b}
\end{eqnarray}
Alternatively, it is also possible to calculate
$J$ from the dc component of the current flowing trough an arbitrary
bond $\langle l, l+1 \rangle$ of the central tight-binding
chain;
\begin{equation}
J_{l,l+1}(t)= 2 e w_{l,l+1} \mbox{Re}[G^<_{l,l+1}(t,t)].
\label{alter}
\end{equation}
Due to the continuity property, the dc component of (\ref{alter})
is independent of $l$ and coincides with the result obtained from 
(\ref{jdc2b}). 

We consider zero temperature
and the same chemical potential $\mu$ for the two reservoirs, hence 
$f_{\alpha}(\omega)=f(\omega)=\Theta(\mu-\omega)$. 
An interesting representation of $J$ is found expressing it in terms of a 
 {\em transmission } function  $T(\omega)$:
\begin{equation}
J= e \int_{-\infty}^{\infty}  d\omega f(\omega) T(\omega).
\label{trans}
\end{equation}
Evaluating $G^<_{l,l+1}(t,t)$ from (\ref{dyson}) the explicit equation
for the transmission function reads
\begin{eqnarray}
 T(\omega) & = & \frac{w_{l,l+1}}{\pi \tau_0} \int_0^{\tau_0} dt \{|w_L|^2 \rho_L(\omega) \mbox{Im}[
G^R_{l,1}(t,\omega) 
G^A_{1,l+1}(\omega,t)] \nonumber \\
& &  + |w_R|^2 \rho_R(\omega) \mbox{Im}[
G^R_{l,N}(t,\omega) G^A_{N,l+1}(\omega,t)] \}.
\label{tl}
\end{eqnarray}
The above representation of $J$  
exhibits a very
important difference between the current generated in a quantum pump
and the stationary transport caused by reservoirs at
 different potentials. In the latter situation, the current is
expressed as $J=e \int d\omega T(\omega) [f_L(\omega)-f_R(\omega)]$.
Namely, the spectral contribution of $T(\omega)$ to the net current 
corresponds to states with energies
 between the two different chemical potentials. Instead, in  the pump
 {\em all } the states bellow the 
Fermi energy of the reservoirs contribute to the net current. 
Another important difference between these two kinds of transport
mechanisms is the origin of the direction of the current.
In stationary transport, $T(\omega)$ is a possitive-defined function,
which is interpreted as the probability of tunneling, while the sign of
$J$ is determined by the bias through $f_L(\omega)-f_R(\omega)$. Instead, in time-
dependent transport without static bias, the transmission function 
(\ref{tl}) can be either possitive or negative and can also change sign as a   
function of $\omega$. This function can be interpreted as the difference
between the probability of tunneling from left to right and the
probability of tunneling from right to  left.

The next sections are devoted to evaluate explicitly $T(\omega)$ and $J$ for
the particular cases of one and two time-dependent potentials and to analyze
in detail their behavior.

\section{One harmonically time-dependent potential.}
\subsection{General considerations.}
The treatment exposed in the previous section simplifies considerably for the
case of 
only
one harmonically time-dependent potential ($M=1$). 
The  
Hamiltonian for the central system reads:
\begin{equation}
H_C(t)=[ V \cos(\Omega_0 t) -\varepsilon_1 ]c_1^\dagger c_1.
\label{dot}
\end{equation}
The reservoirs $L$ and $R$ are placed at the left and the
right, respectively, of the pumping center, as explained in the previous subsection. 
Our aim is the calculation of the net
current.
Using the expansion (\ref{four2}), the dc components of the currents flowing
from the reservoirs towards the central  site read
\begin{eqnarray}
& & J^{dc}_{\alpha}= e |w_{\alpha}|^2\int_{-\infty}^{\infty} \frac{d \omega}{2 \pi} f(\omega) \{
- 2 \mbox{Im}[{\cal G}_{1,1}(0,\omega)] \rho_{\alpha}(\omega) \nonumber \\
& & -\sum_k \rho_{\alpha}(\omega -  k \Omega_0) \Gamma(\omega) 
|{\cal G}_{1,1}(k,\omega) |^2 \},
\label{curr2}
\end{eqnarray}
being $\Gamma(\omega) =|w_L|^2 \rho_L(\omega)+ |w_R|^2 \rho_R(\omega) $.
It is clear that a necessary condition
for a non-vanishing $J^{dc}_{\alpha}$ is that the two terms of (\ref{curr2})
do not cancel one another.

Further insight is gained by using the representation of the current in terms
of the transmission function.
In the present case, it is possible to solve recursively the set (\ref{dyret})
to
calculate the retarded Green function. This
procedure is summarized in Appendix A. 
After replacing the expressions (\ref{rel}) in (\ref{jdc2b}),
with $N \equiv 1$,  it is found
\begin{eqnarray}
& & T(\omega) = \frac{1}{\pi}  |w_L|^2 |w_R|^2  \sum_{k=-\infty}^{\infty}
|{\cal G}_{1,1}(k,\omega) |^2 \nonumber \\
& & \times 
  \{ \rho_L(\omega) \rho_R(\omega-k \Omega_0)- \rho_R(\omega)
\rho_L(\omega-k \Omega_0) \},
\label{curr3}
\end{eqnarray}
where it becomes clear that $J=0$ when (i) the system is
symmetric under spacial inversion centered at
the point where the pumping potential is applied, such that 
$\rho_L(\omega)=\rho_R(\omega)$, and (ii) the environment of the pumping 
point can be
described by flat and approximately constant
densities of states $\rho_{\alpha}(\omega)$.
This expression for $T(\omega)$ resembles the mechanism of photon-assisted
tunneling. The resulting equation for the current has a similar form as that
obtained within the framework of the scattering matrix formalism, identifying 
the square of the scattering matrix in the Floquet formalism with the function
$ |{\cal G}_{1,1}(k,\omega) |^2 
\rho_L(\omega) \rho_R(\omega-k \Omega_0)$ \cite{mosk}.
The current can be viewed as the result of processes where
electrons leave  the reservoir $\alpha$ with probability
$\rho_{\alpha}(\omega)$, interact with the pumping center loosing or
gaining $k$ energy quanta $\Omega_0$ with probability
$\propto |{\cal G}_{1,1}(k,\omega) |^2$, and exit to the opposite 
reservoir $\beta$ with probability $\rho_{\beta}(\omega-k\Omega_0)$.
The sign of the net current is completely determined by the structure of
the functions $\rho_{\alpha}(\omega)$. Note that $T(\omega)$ may
change sign as a function of $\omega$. This means that electrons with
different energies can flow in different directions and it is the 
sum of all these contributions what determines the sign of the net current.

For small 
$\Omega_0$ and $V$, only a few modes $k$ contribute. 
It is natural to associate such a situation with
 the idea of {\em adiabatic} pumping. More precisely,
the concept of adiabatic pumping applies to the regime
where the characteristic time scale for an electron to travel across
the pump (proportional to the inverse of the width of the spectral peaks of ${\cal G}(k,\omega)$
as a function of $\omega$) is much smaller than $\Omega_0$. Adopting that
definition, we see that it is possible in this case to have a finite $J$ even
in the adiabatic regime. The key is a high hybridization $w_R, w_L$ of the
pumping center with the environment, in order to 
 allow for wide peaks in ${\cal G}(k,\omega)$, meaning a short 
life of the electrons at the pumping center. For low pumping frequencies,
(\ref{curr3}) can be expanded in powers 
of $\Omega_0$ and it is found $T(\omega) \propto \Omega_0$.

A final interesting remark is that $J \propto |w_L|^2 |w_R|^2$. This kind of
behavior has already  been  found in molecular ratchets  pumped by a laser field
\cite{leh}, which are modeled on the basis of tight-binding Hamiltonians with
asymmetric energy profiles and sincronic pumping centers.

\subsection{Example}
To illustrate the discussion of the previous subsection we 
show some results of local pumping in 
a symmetric double barrier structure with the pumping potential
 acting at one of the barriers.
The ``unperturbed'' structure has spacial inversion
symmetry with respect to the center, which is broken by the effect of the
time-dependent voltage. 

\begin{figure}
\includegraphics[width=8cm,clip]{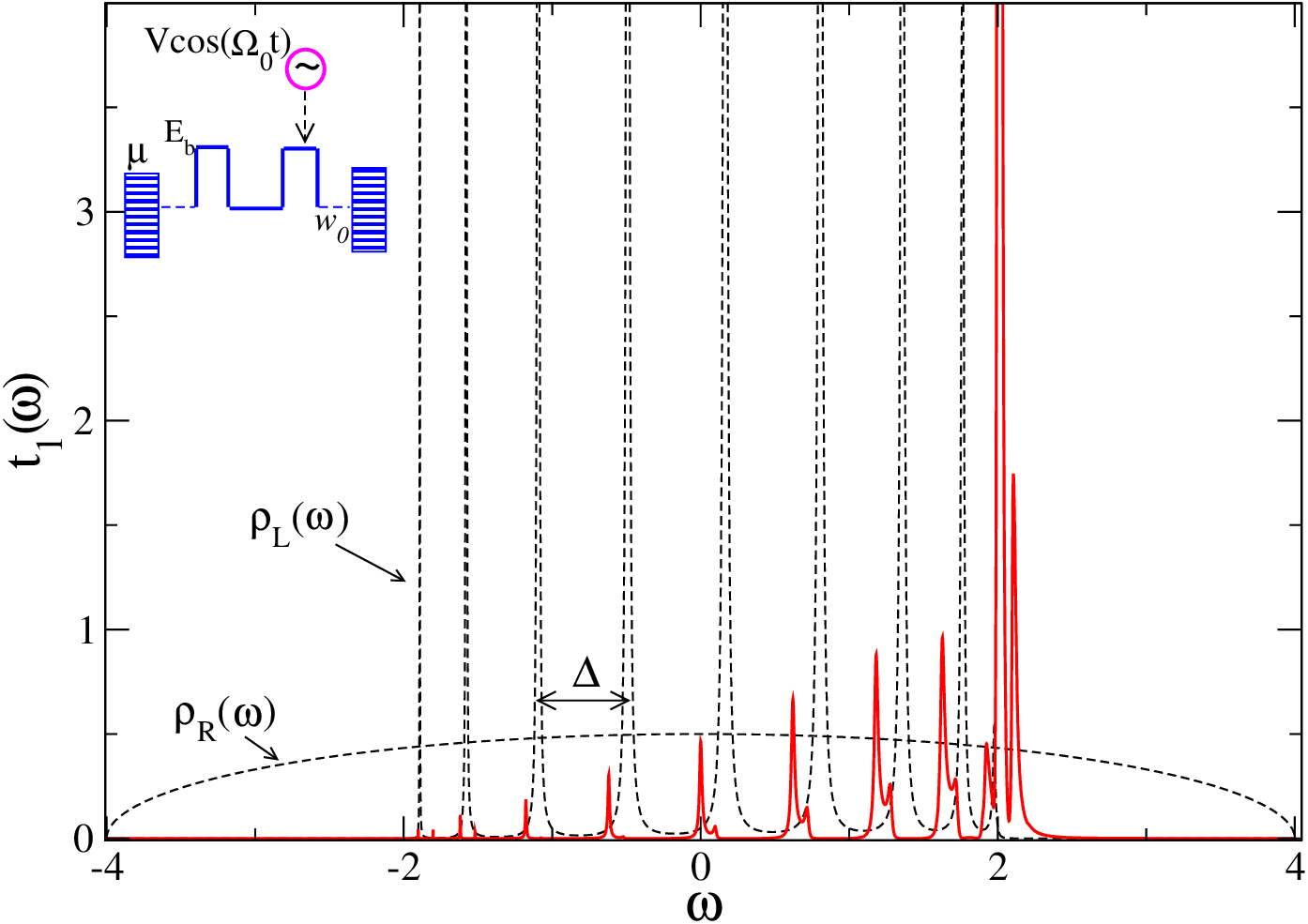}
\caption{(Color online) Solid red line: the function $t_1(\omega)=|{\cal
    G}(1,\omega)|^2$ for $V=0.1$, $\Omega_0=0.1$. 
Dashed black lines: The densities of states $\rho_L(\omega)$ and
$\rho_R(\omega)$  of
 the left and right  reservoirs, respectively.
$\Delta$ denotes the energy difference between two consecutive
energy levels in the double barrier structure.
A scheme of the device is indicated in the upper left corner.
Other parameters are $E_b=1$,  $|w_0|^2=0.1$, $W=4$, $N_L=9$. }

\label{fig1}
\end{figure}

A scheme of the device is shown in Fig. \ref{fig1}.
For sake of clarity we write down the model Hamiltonian:
\begin{eqnarray}
& & H =  H_{leads} + H_{cont} -w \sum_{l=-N_L}^{0}  c^{\dagger}_l c_{l+1} + Hc + \nonumber \\
& &
\sum_{l=-N_L}^{1} \varepsilon_l  c^{\dagger}_l c_{l}  + V \cos(\Omega_0 t)  c_1^\dagger c_1,
\end{eqnarray}
where $\varepsilon_l$ defines the profile corresponding to the
barriers. We consider a two-barrier
structure of height $E_b$:  
$\varepsilon_{-N_L}=\varepsilon_{1 }= E_b$ and 
$\varepsilon_l=0, l\neq -N_L,1$.
We denote with $H_{leads}$ the Hamiltonians of 
two semi-infinite chains which behave as macroscopic reservoirs and represent
two external leads connected to the central device.  These parts are
pictorially represented by boxes with stripes in the scheme of
Fig. \ref{fig1} and we describe them by semicircular densities of states
with bandwidth $W$,
\begin{equation}
\rho^0(\omega)= 4 \sqrt{1-\omega^2/W^2} \Theta(W - \omega).
\label{circ}
\end{equation}
The term
$H_{cont}$ describes the hopping between the semi-infinite leads and
the central structure and has the form of the last two terms of the 
Hamiltonian (\ref{ham1d}), with hoping parameter $w_0$.

In the notation
of Eqs. (\ref{dot}) and (\ref{ham1d}), the right reservoir corresponds to the
right lead, while the left reservoir corresponds to the left lead plus the 
double barrier structure with the exception of the point $l=1$, where the pumping
voltage is applied. The ensuing degrees of freedom can be easily integrated out defining
the following  retarded Green functions for the left and right reservoirs:
\begin{equation}
g^R_L(\omega)=g^0(\omega)=\int_{-\infty}^{\infty}\frac{d\omega'}{2 \pi}
\frac{\rho^0(\omega')}{\omega - \omega'+ i\eta},
\end{equation}
with $\eta=0^+$,
and 
\begin{equation}
g^R_R(\omega)=\displaystyle\frac{1}
{\omega - \varepsilon_2 - \displaystyle\frac{w^2}
{\omega - \varepsilon_3 - \ldots - \displaystyle\frac{w^2}
{\omega - \varepsilon_{N_R}  - {w_0}^2 g^0(\omega) 
}
}
},
\end{equation}
being the densities of states of the reservoirs
$\rho_{\alpha}(\omega)=- 2 \mbox{Im}[g^R_{\alpha}(\omega)]$.
The hoppings between the pumping center and the left and right reservoirs
are $w_L=w$  and
$w_R=w_0$, respectively.

Fig. \ref{fig1} illustrates the behavior of the  function
$t_k(\omega)=|{\cal G}_{1,1}(k,\omega) |^2$, related to the
probability for an incoming electron with energy $\omega$ to loose 
$k$ energy quanta $\Omega_0$ for selected 
parameters, along with the densities of states  of the
left and right reservoirs. All the energies are written in units of the hopping parameter
$w$, which is set to $w=1$. 
All currents are expressed in units where $e=1$.
The results shown in the figure correspond to 
a weak pumping amplitude ($V=0.1$), where
only the contribution $k=1$ is sizable. For larger $V$, higher modes come also
into play.

\begin{figure}
\includegraphics[width=8cm,clip]{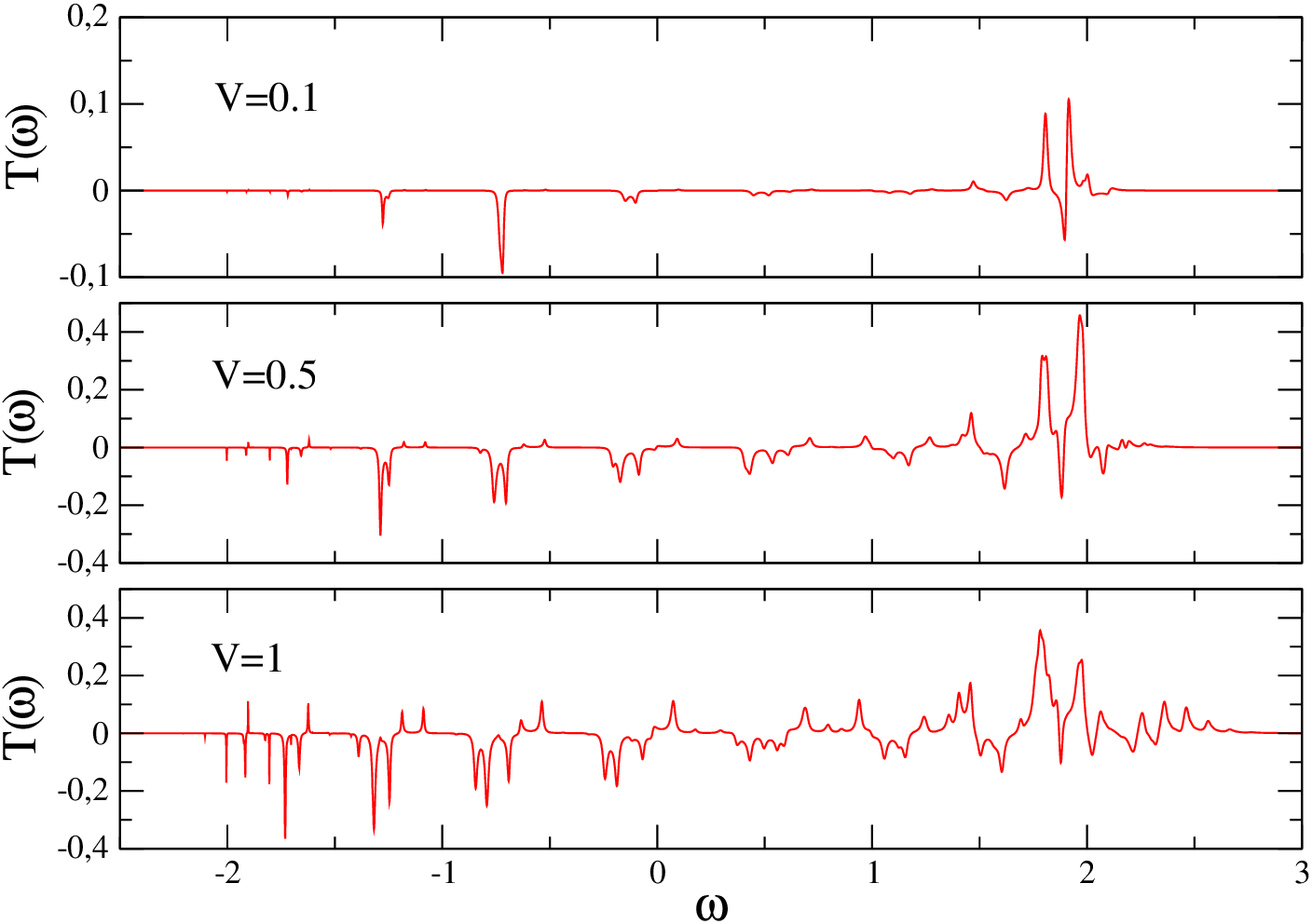}
\caption{(Color online) The transmission function $T(\omega)$
for $\Omega_0=0.1$ and $V=0.1,0.5,1$ (top to bottom).
Other parameters are as in Fig. \ref{fig1}.}
\label{fig2}
\end{figure}

The resulting structure of the transmission function $T(\omega)$ is shown
in Fig. \ref{fig2} for a low frequency $\Omega_0$ and different amplitudes
$V$. The important feature to note is that for high $\omega$ and large
pumping amplitude, this function can experiment several changes of sign. This
indicates that electrons with different energies may flow along different directions.
The direction of the total 
direct current being the cumulative sum of all these contributions.  

\begin{figure}
\includegraphics[width=8cm,clip]{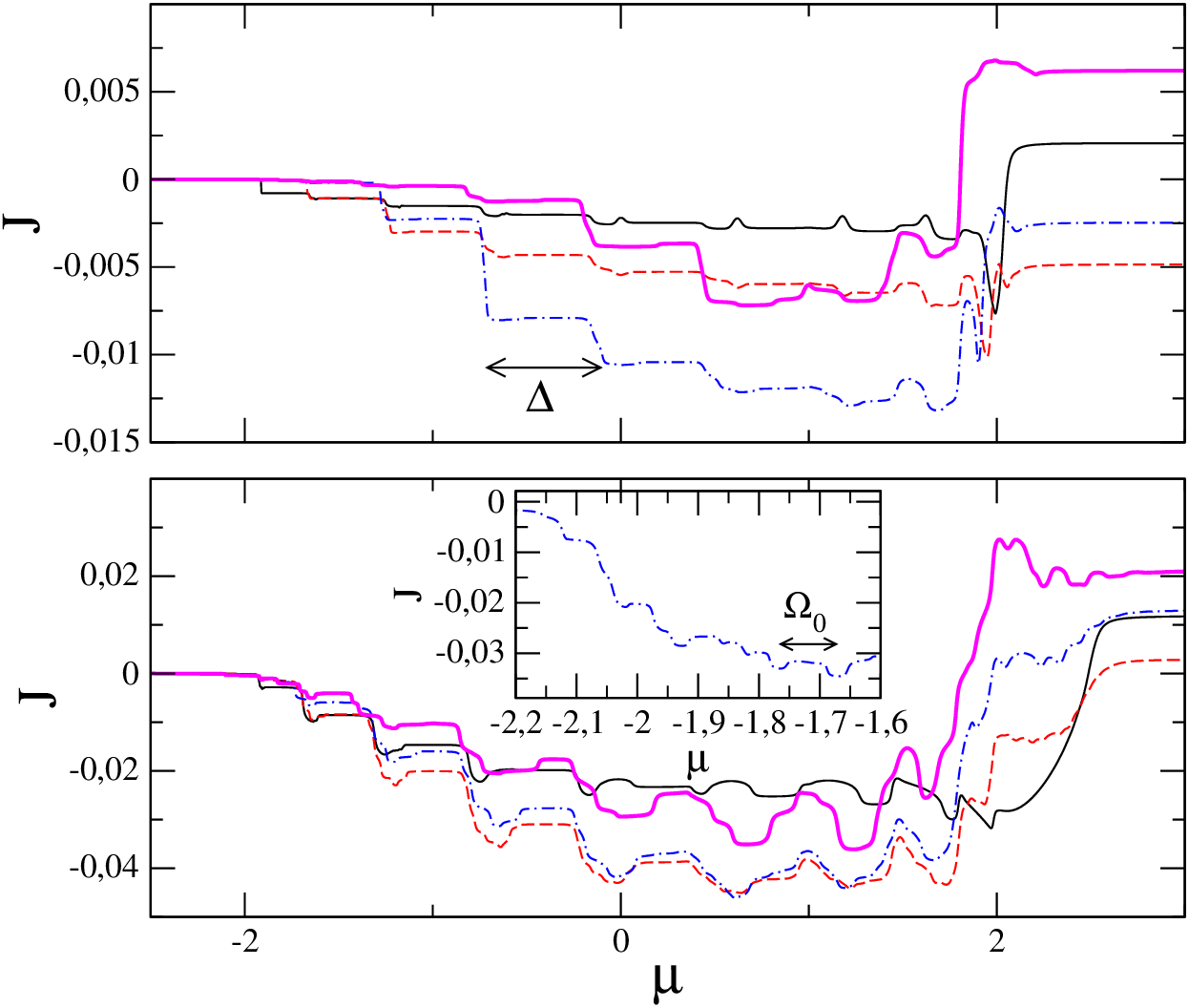}
\caption{(Color online) The dc current $J$ as a function
of the chemical potential of the electrodes $\mu$. Top and
bottom panels corresponds to pumping voltages $V=0.2$ and
$V=1$, respectively. Plots in thin solid
black, dashed red, dot-dashed blue and thick solid magenta lines
correspond to $\Omega_0=0.01,0.05,0.1,0.2$, respectively.
The spacing between two consecutive levels is indicated with $\Delta$.
The inset shows details for  $V=2$ and $\Omega_0=0.1$.
Other parameters are as in Fig. \ref{fig1}. 
}
\label{fig3}
\end{figure}
Some results on the behavior of the net current $J$ as a function of the
chemical potential $\mu$ of the reservoirs are shown in Fig. \ref{fig3}.
The structure of jumps and plateaus observed in the figure follows the
patterns of resonances related to the peaks of $t_k(\omega)$ and
$\rho_L(\omega)$. The energy interval between two consecutive
jumps in $J$ is roughly the difference of energy between two energy levels
$\Delta$. 
For strong pumping amplitude, many $k$
modes come into play in the transport process. The transmission function develops
a richer structure where the spectral weight associated to the free electronic levels
split into several side peaks separated in $\sim \Omega_0$ (see bottom panel of Fig.
\ref{fig2}). This is translated in peaks at the edges of the plateaus shown
in $J$ as a function of $\mu$ as shown in the bottom panel of Fig. \ref{fig3}.
Details of the structure related to $\Omega_0$ is shown in the inset.

For $\mu < 2$, the chemical potential lies below the highest 
resonance of $\rho_L(\omega)$ and 
 the behavior of $T(\omega)$ is consistent with
a preference in the flow right to left, while
for  higher $\mu$, the flow can take place in the opposite direction. 
Details depend  on the frequency and amplitude of the pump, as well as on
the degree of coupling between the double-barrier structure and the
macroscopic reservoirs.

The behavior of the sign of $J$ for $\mu < 2$ is roughly the one expected from
an intuitive adiabatic description. The picture that emerges is as follows:
during the  part of the pumping cycle, where the potential
decreases,  the tunneling from the right macroscopic
reservoir into the quantum well is favored, while during the remaining part of the cycle,
the total potential at the right barrier increases and the electrons
accumulated
in the well are
pushed from the right to the left.
Remarkably, within this range of $\mu$, the sign of $J$ inferred from such a simple
description remains unchanged for high $\Omega_0$ and strong $V$, where the
adiabatic picture is not expected to be valid. 
The situation for higher $\mu$ is much less clear and strongly depends on
the particular values of the pumping parameters.

To finalize, we want to remark that, due to the symmetry of the problem,
the situation where the pumping potential is applied at the left barrier,
would result in exactly the same behavior of $J$, but with the opposite 
sign. As discussed in the previous subsection, a vanishing $J$ would be
obtained if the pumping potential were applied exactly at the center
of the well between the two barriers.

\section{Two harmonically time-dependent potentials in a double barrier structure.}
\subsection{General considerations}
We now go back to Hamiltonian
(\ref{ham1d}) and 
consider the following Hamiltonian for the central region:
\begin{eqnarray}
& & H_C (t)  =  
V \cos(\Omega_0 t + \delta ) c_1^\dagger c_1 
 +\sum_{l=1}^N \varepsilon_l c^\dagger_l c_l \nonumber \\
& &  - w \sum_{l=1}^N (c_l^\dagger c_{l+1} +H.c) +
V \cos(\Omega_0 t ) c_N^\dagger c_N,
\label{ham2}
\end{eqnarray}
with the profile $\varepsilon_1=\varepsilon_N=E_b$, $\varepsilon_l=0,
l=2,\ldots,N-1$,
defining a double barrier structure. This arrangement is similar to
the one of the experimental setup of Ref. \cite{swi}, where two ac potentials
with a phase-lag
are applied at the walls confining a quantum dot.

In the interesting case of reservoirs with wide bands, it is
possible to find an explicit  relationship between the current and the
local spectral functions at the points where the pumping potentials are applied.
The wide band limit corresponds to approximately constant
densities of states $\rho_{\alpha}(\omega) \sim \rho_{\alpha}$, such that
\begin{equation}
 \int_{-\infty}^tG^<_{i,i}(t,t_1) g^A_{\alpha}(t_1-t)   \sim
 i G^<_{i,i}(t,t) \rho_{\alpha},
\label{wide}
\end{equation}
with $i=1$ for $\alpha=L$ and $i=N$ for $\alpha=R$.
The 
expression (\ref{jdc2b}) for the dc current leads to 
\begin{eqnarray}
& & T(\omega)= |w_{L}|^2 [
\rho_{L}(\omega) [\rho^{dc}_{1}(\omega) - 
\overline{\rho}_1(\omega)] \nonumber \\
& & -|w_{R}|^2
\rho_{R}(\omega) [ \rho^{dc}_{N}(\omega) - 
\overline{\rho}_N(\omega)],
\label{curtran}
\end{eqnarray} 
where we have defined the dc component  of the
 generalized densities of states at the sites where
the time-dependent potentials act:
\begin{equation}
\rho^{dc}_{i}(\omega) = - \frac{1}{\tau_0} \int_0^{\tau_0} dt
2 \mbox{Im}[G^R_{i,i}(t,\omega)],
\end{equation}
and the dc components of the spectral densities of occupation at those
sites, $\overline{\rho}_i(\omega)$: 
\begin{eqnarray}
& & \overline{\rho}_i(\omega) = \frac{1}{\tau_0} 
\int_0^{\tau_0} dt \{ |G^R_{i,1}(t,\omega)|^2 |w_L|^2 \rho_L(\omega) \nonumber \\
& &+ |G^R_{i,N}(t,\omega)|^2 |w_R|^2 \rho_R(\omega),
\nonumber \\
& &  n_i = -i \frac{1}{\tau_0} \int_0^{\tau_0} dt 
G^<_{i,i}(t,t) =  \int \frac{d\omega}{2 \pi}  
f(\omega) \overline{\rho}_i(\omega),
\label{den}
\end{eqnarray}
being $n_i$  the density of particles at the site $i$.
In an equilibrium system, these two spectral functions coincide, i.e.,
$\overline{\rho}_i(\omega) \equiv \rho^{dc}_{i}(\omega)$
and are positive defined functions, but in a time dependent problem these
two functions differ in general. While $\overline{\rho}_i(\omega)$ is a positive defined
function, in a time-dependent problem $\rho_i(t,\omega)$ may change sign as
a function of $\omega$. 
Actually, it is clear from Eq. (\ref{curtran}) that 
the violation of the  equivalence between these two spectral functions,
is at the heart of the
existence of a non-vanishing  dc current. Another necessary condition
is the breaking of left-right symmetry. 
This can be accomplished statically by, for example, considering
$\rho_L (\omega) \neq \rho_R (\omega)$ or $w_L \neq w_R $, but also dynamically, by
recourse to a finite phase lag $\delta \neq 0 $ in the Hamiltonian 
(\ref{ham2}). In what follows, we focus on the latter case.

\subsection{Results}
 \begin{figure}
\includegraphics[width=8cm,clip]{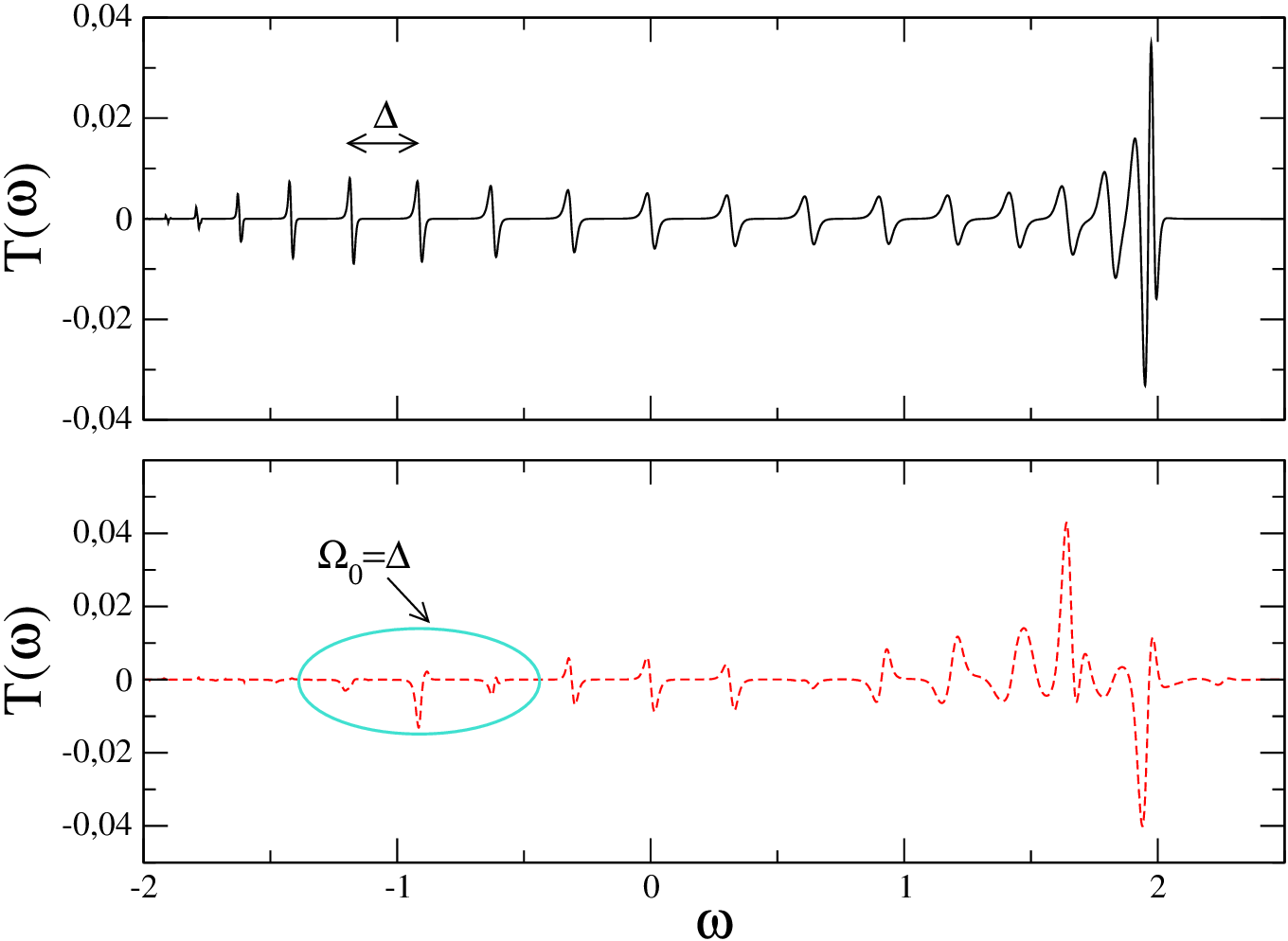}
\caption{(Color online) The transmission function $T(\omega)$
for a pumping amplitude $V=0.2$ a phase lag
$\delta=\pi/2$ and frequencies $\Omega_0=0.01,
0.3$ (top and bottom panels).
The ellipse in the lower panel encloses a region where resonance is achieved.
 Other parameters are
$E_b=1$, $|w_0|^2=0.5$, $W=4$ and $N=20$.}
\label{fig4}
\end{figure}
We consider the 
symmetric array of two barriers described in the previous subsection, which are
placed between 
two identical reservoirs 
with a large
bandwidth  
$\rho_L(\omega)=\rho_R(\omega)= \rho^0(\omega)$ with the form defined in
(\ref{circ}),
and $|w_L|^2= |w_R|^2=|w_0|^2$.
Due to the symmetry of the setup,  $\rho^{dc}_1(\omega) = \rho^{dc}_N(\omega)$, while $\delta$
causes differences in the densities of occupation $\overline{\rho}_1(\omega)$ and 
$\overline{\rho}_N(\omega)$. In this way, the transmission function (\ref{curtran})
simplifies to 
\begin{equation}
T(\omega)= |w_0|^2 \rho^0(\omega) [\overline{\rho}_N(\omega)- 
\overline{\rho}_1(\omega)],
\label{tsim} 
\end{equation}
and it 
 is found that 
the current behaves like $J \propto (n_N-n_1)$. This is very suggestive, since it is
natural to associate a difference in the local density of particles with a
static potential drop. In the present case, we can imagine that the pumping
with a
phase lag induces an effective potential drop $V_{eff}$  between the sites 
$1$ and $N$,
 which causes a difference in the particle population at these two sites, 
and  a current  $J \propto V_{eff}$. 

An important situation corresponds to the case of weak pumping amplitude $V$.
A perturbative solution of the set (\ref{dyret}) to the lowest order in $V$ 
leads to 
\begin{eqnarray}
 G^R_{m,n}(t,\omega)  & \sim & 
G^0_{m,n}(\omega)+\frac{V}{2}   e^{i (\delta+\Omega_0 t)}
G^0_{m,1}(\omega-\Omega_0)G^0_{1,n}(\omega) \nonumber \\ 
& & +
\frac{V}{2}  e^{i \Omega_0 t} G^0_{m,N}(\omega-\Omega_0) 
G^0_{N,n}(\omega) + \nonumber \\
& & \frac{V}{2}   e^{-i (\delta+\Omega_0 t)}
G^0_{m,1}(\omega+\Omega_0)G^0_{1,n}(\omega) + \nonumber \\
& &
\frac{V}{2}   e^{-i \Omega_0 t} G^0_{m,N}(\omega+\Omega_0) 
G^0_{N,n}(\omega).
\label{dyretper}
\end{eqnarray}
Using these expressions for the Green functions,
making use of the fact that for the symmetric device 
$G^0_{11}(\omega)=G^0_{NN}(\omega)$ and $G^0_{1N}(\omega)=G^0_{N1}(\omega)$
and keeping terms to the lowest non-vanishing order in  $V$, we find
\begin{eqnarray}
 T(\omega) & \sim & 2 V^2 |w_0|^4  \sin(\delta) [\rho^0(\omega)]^2
\mbox{Re}\{G^0_{11}(\omega) [G^0_{1N}(\omega)]^* \} \nonumber \\
& & \times \mbox{Im}\{ G^0_{11}(\omega + \Omega_0)
[G^0_{1N}(\omega+\Omega_0)]^* 
\nonumber \\
& &
-  G^0_{11}(\omega - \Omega_0) [G^0_{1N}(\omega - \Omega_0)]^* \}.
\label{tap}
\end{eqnarray}

\begin{figure}
\includegraphics[width=8cm,clip]{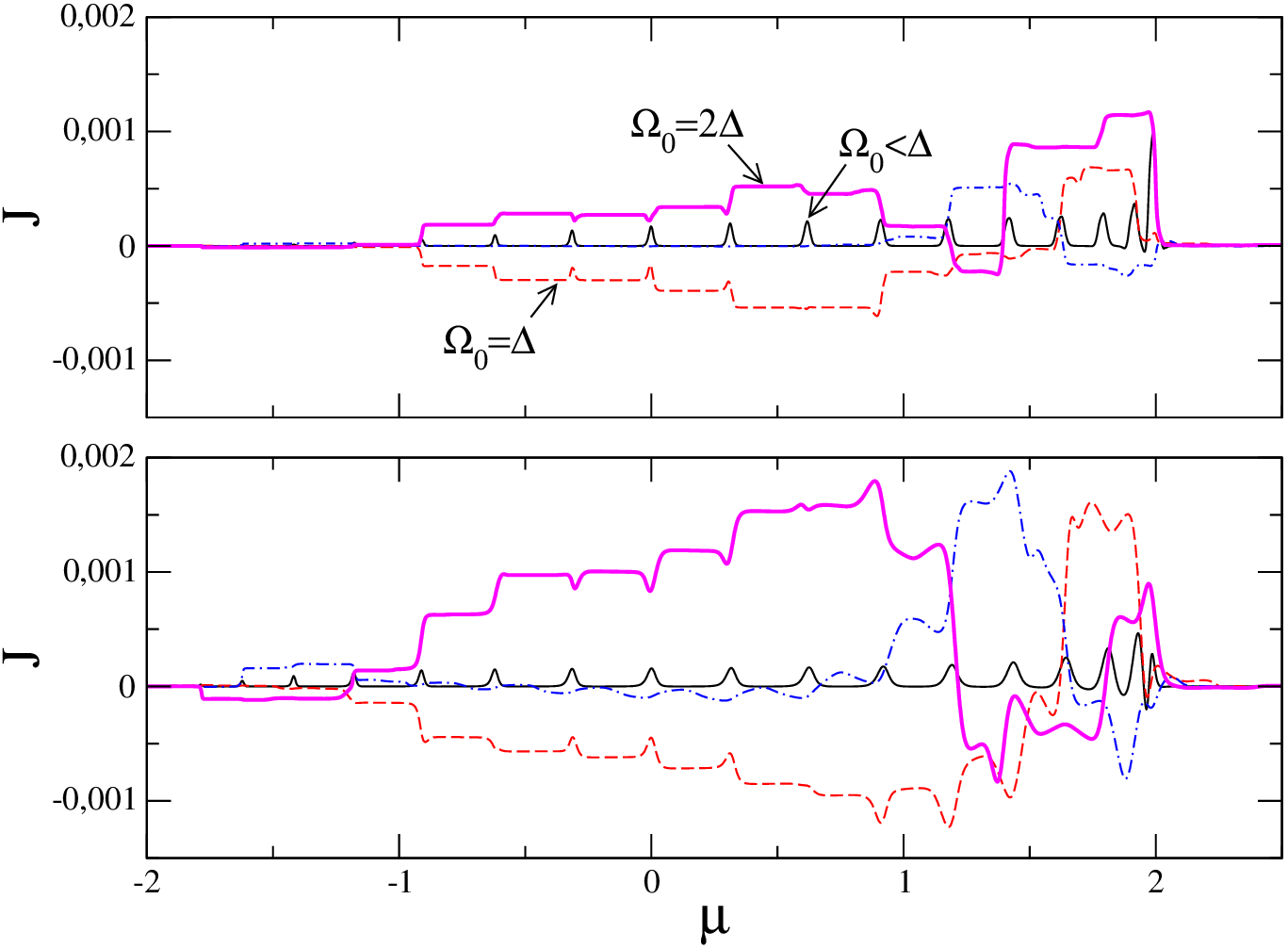}
\caption{(Color online) The dc current $J$ as a function
of the chemical potential of the electrodes $\mu$. Top and
bottom panels corresponds to couplings to the macroscopic reservoirs
$|w_0|^2=0.1$ and $|w_0|^2=0.5$, respectively. The pumping amplitude
is $V=0.2$. Plots in thin solid
black, dashed red, dot-dashed blue and thick solid magenta lines
correspond to $\Omega_0=0.01,0.3,0.45,0.6$, respectively.
The spacing between two consecutive levels is referred to as $\Delta$.
Other parameters are as in Fig. \ref{fig4}. 
}
\label{fig5}
\end{figure}

In order to have a better insight on the behavior of the function
$T(\omega)$, 
 we present some results in Fig. \ref{fig4} for a phase lag $\delta = \pi/2$. 
 In the figure,
we show the exact $T(\omega)$, calculated from eq. 
(\ref{tl}). As the bandwidth  of the macroscopic leads ($2W=8$) is large compared 
to the spectral width of the central piece ($\sim 4$), the wide-band description for the 
reservoirs applies and we have checked that $T(\omega)$ is well reproduced
by equation (\ref{tsim}).
The upper panel of Fig.  \ref{fig4} corresponds to small pumping amplitude $V$,
small pumping frequency $\Omega_0$ and $\delta=\pi/2$. The transmission function shows 
 a peak-antipeak pattern with a separation $\Delta$ that coincides with the
energy distance between two consecutive energy levels of the central structure. 
The peak-antipeak behavior of $T(\omega)$ indicates that, as the
chemical potential increases and covers a level, electrons with  lower energies
within the linewidth
are allowed to travel from the left to the right while the ones with  higher energy
travel in the opposite direction. 
The corresponding
behavior of the dc current as a function of $\mu$ is shown in thin black
line in the upper panel of Fig. \ref{fig5} and it consists in a succession of small peaks suggesting
that as the chemical potential is increased covering an energy level of the
central system, a conduction channel is enabled and a net current flows
between the two reservoirs. The sign of the current is consistent with the
one expected from intuitive considerations for this value of the phase lag
on the basis of the following adiabatic picture:
the first barrier lowers its effective potential during the part of
the cycle where the second potential grows, favoring the incoming of
electrons from the left reservoir inside the well. In the other
part of the cycle, the voltage of the second barrier gets lower, helping
the electrons to tunnel from the well towards the right reservoir. 

An interesting situation takes place when the frequency is resonant, i.e. 
$\Omega_0=\Delta$. In this case, two neighboring electronic levels of the
central device
are expected to be mixed by the pumping potentials.
The region marked in the lower panel of Fig. \ref{fig4}
satisfies the resonant condition and the function $T(\omega)$ consists of
a sequence of antipeaks. The corresponding current is shown in dashed
red lines in the upper panel of Fig. \ref{fig5} and exhibits  plateaus within this region.
The sign corresponds to a net electronic flow from the right to the left
suggesting that  the quantum interference due to the mixing of levels
causes a sign reversal in comparison to the situation observed
for pumping frequencies $\Omega_0 < \Delta$. Also note in Fig. \ref{fig5} that when the resonant 
condition is $\Omega_0=2 \Delta$, the current recovers the sign of the
situation $\Omega_0 < \Delta$. 
For weak pumping amplitude $V$,  a further
examination on the origin of this interference is possible on the basis of the
approximate solution given in Appendix B. The analysis presented there
indicates that, at resonance, the effective phase lag between the two potentials is 
$\delta + j \pi$, being $j-1$ the number of energy levels of the structure
 between the two interfering ones.
On the other hand, due to the symmetry of the problem, $J(\delta+j\pi)= (-1)^j
J(\delta)$, which complements the argument to explain why a shift in the phase
lag may cause a change in the sign of the net current.
The lower
panel of Fig. \ref{fig5} gives an idea of the effect of the changes in the
linewidth of the central piece introduced through an stronger coupling to the
reservoirs. As expected, these effects are more important at resonant
frequencies since they help in the mixing of the electronic levels.
For high energies, close to the upper edge of the spectrum ($\omega\sim 2$), 
the energy levels are closer ($\Delta$ is smaller). Therefore, the resonant
frequencies strongly differ from the ones at lower energies and sign reversals of
$J$ are observed as a function of $\mu$.

\begin{figure}
\includegraphics[width=8cm,clip]{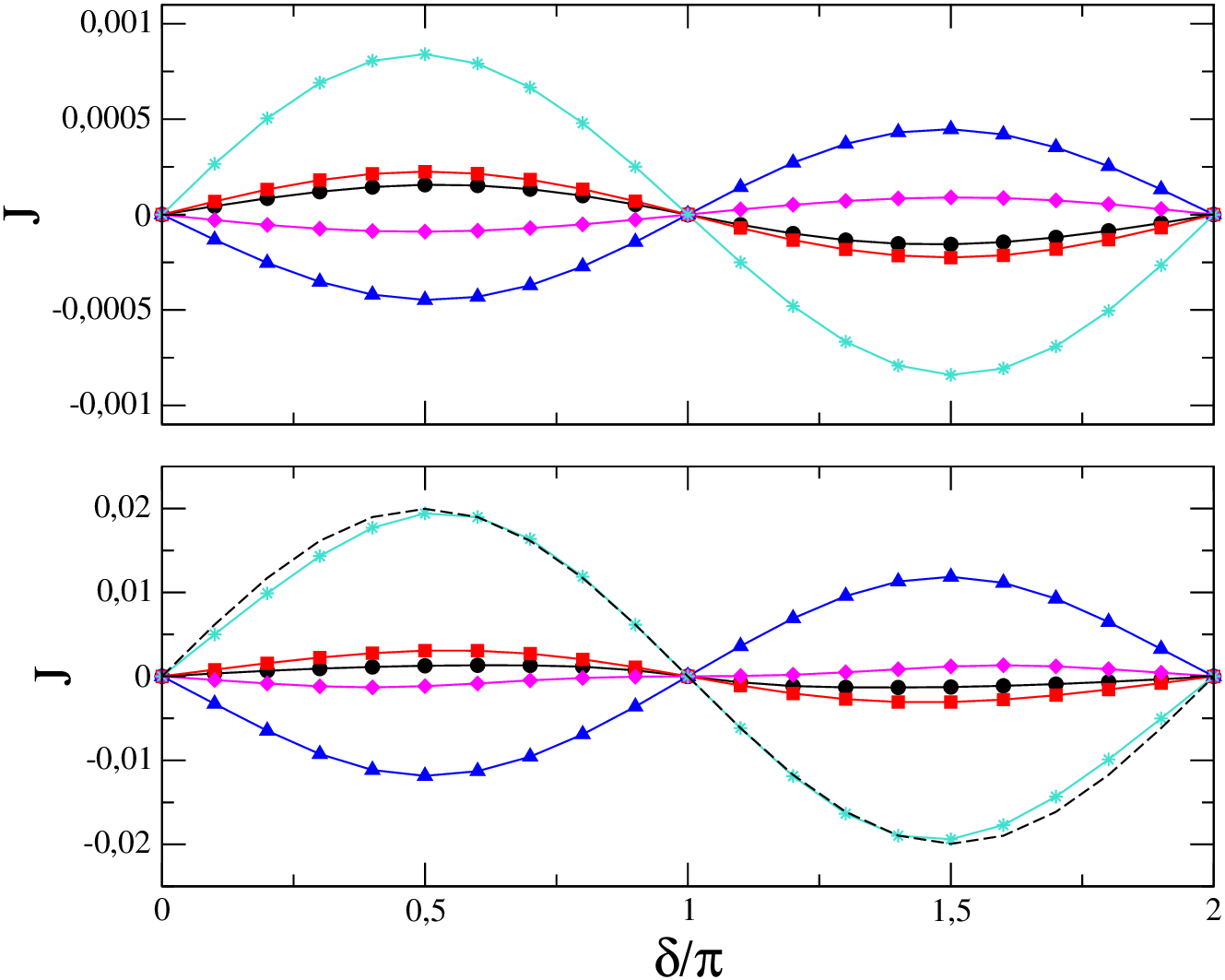}
\caption{(Color online) The dc current $J$ as a function
of the phase lag $\delta$ for $\mu=0$
and $|w_0|^2=0.5$. Upper and lower panels correspond to  pumping amplitudes
$V=0.2$ and $V=1$, respectively. Circles, squares, triangles, diamonds and
stars correspond to 
 $\Omega_0=0.01,0.05,0.3,0.45,0.6$, respectively. The function
$J=0.0185 \sin(\delta)$ is indicated in dashed lines in the lower panel.
Other parameters are as in Fig. \ref{fig4}. 
}
\label{fig6}
\end{figure}

The behavior of $J$ as a function of the phase lag is shown in Fig. \ref{fig6}
for a selected value of the chemical potential $\mu$.
 All the plots of the upper panel correspond to a small pumping amplitude
and can be 
fitted by a function $\propto \sin(\delta)$, in full agreement
with eq. (\ref{tap}). 
The lower panel corresponds to a higher $V$ and deviations from this behavior 
are found. For example, the plot in dashed lines corresponds to a function
of the form
$J =A \sin (\delta)$ and fits the behavior of $J$ for $\Omega_0=0.6$ only
in a neighborhood of $\delta=\pi$. Similar deviations are observed for other
pumping
frequencies.

\begin{figure}
\includegraphics[width=8cm,clip]{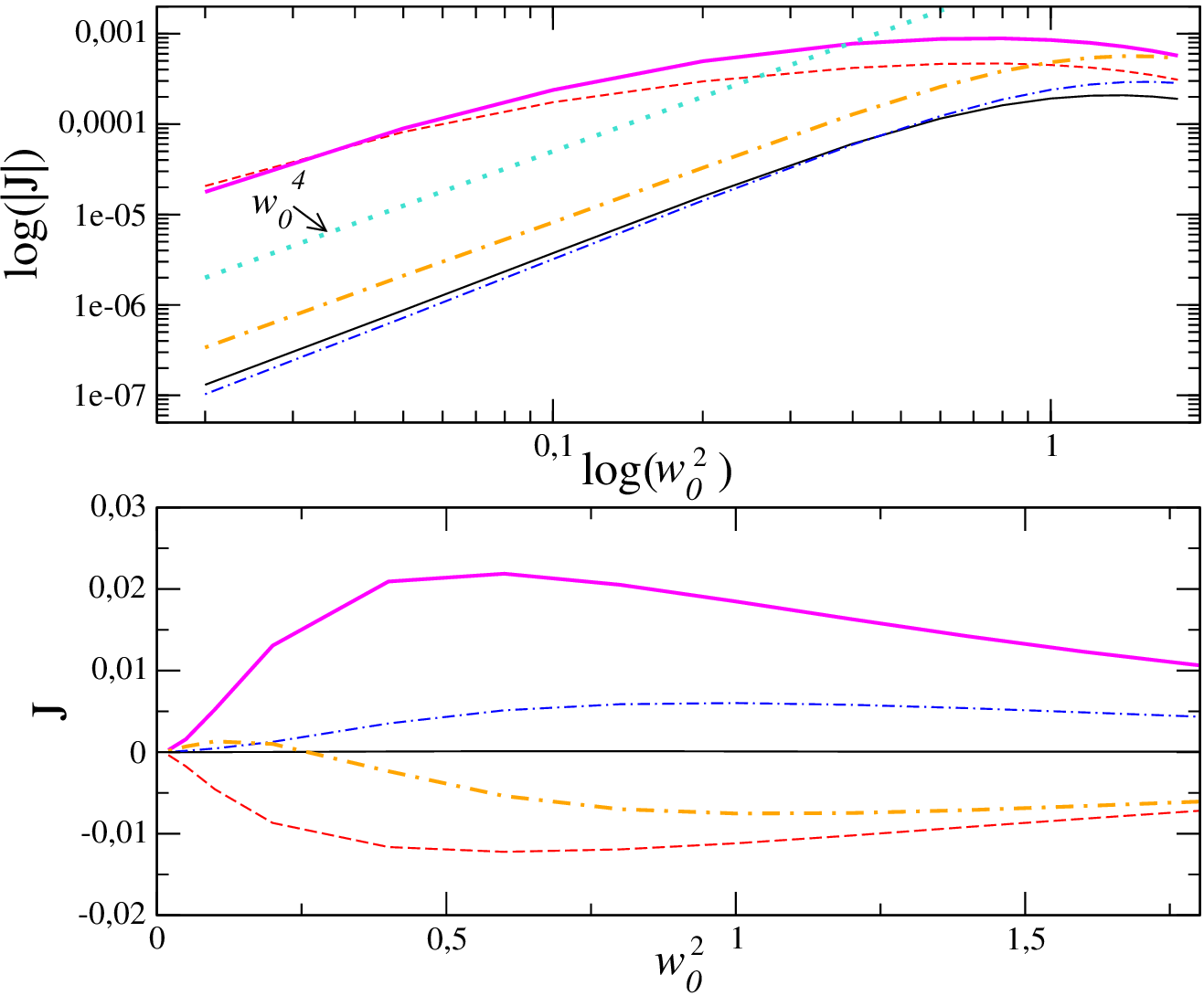}
\caption{(Color online) The dc current $J$ as a function
of the contact parameter 
$ |w_0|^2$ for $\delta=\pi/2$.
The upper panel is plotted in log-log scale and 
corresponds to a pumping amplitude $V=0.2$ and chemical potential $\mu=0$.
The reference slope corresponding to $J \propto |w_0|^4$ is plotted in light
blue
dotted
line. The lower panel corresponds to 
 $\mu=1$ and pumping amplitude $V=1$. Plots in thin solid
black, dashed red, dot-dashed blue and thick solid magenta
and thick dashed-dot orange lines
correspond to $\Omega_0=0.1,0.3,0.45,0.6,0.75$, respectively.
Other parameters are as in Fig. \ref{fig4}. 
}
\label{fig7}
\end{figure}
The behavior of $J$ as a function of the coupling 
to the reservoirs $|w_0|^2$ is shown in Fig. \ref{fig7} for fixed
$\mu$, with different frequencies and pumping amplitudes.  
The higher panel corresponds to a low pumping amplitude and it is drawn in
log-log scale in order to observe details for very low
coupling to the reservoirs. 
In most of the cases, the current vanishes as the coupling to the reservoirs 
tends to zero
following a law $J \propto |w_0|^4$ as suggested by (\ref{tap}).  The
corresponding reference slope is indicated in the figure for comparison.
In some cases, slight deviations from this law are observed. One  example is
the plot in red dashed lines shown in the upper panel which corresponds to a
resonant frequency. The origin for such departures should be found in the fact
that the functions $G^0_{ij}(\omega)$ tend to be singular as the coupling to
the reservoirs tends to vanish. The  case of
a larger pumping amplitude is illustrated in the lower panel of
Fig. \ref{fig7}. A notable feature observed in the latter case is
that changes in the strength of the coupling
to the reservoirs may introduce changes in the sign of $J$, like in the case
of the plot corresponding to $\Omega_0=0.75$. This is because the coupling
to the reservoirs contributes to enhance the quantum interference between
energy levels. For very small $|w_0|^2$, the behavior is similar to the one
observed for weak $V$, illustrated in the higher panel.

\begin{figure}
\includegraphics[width=8cm,clip]{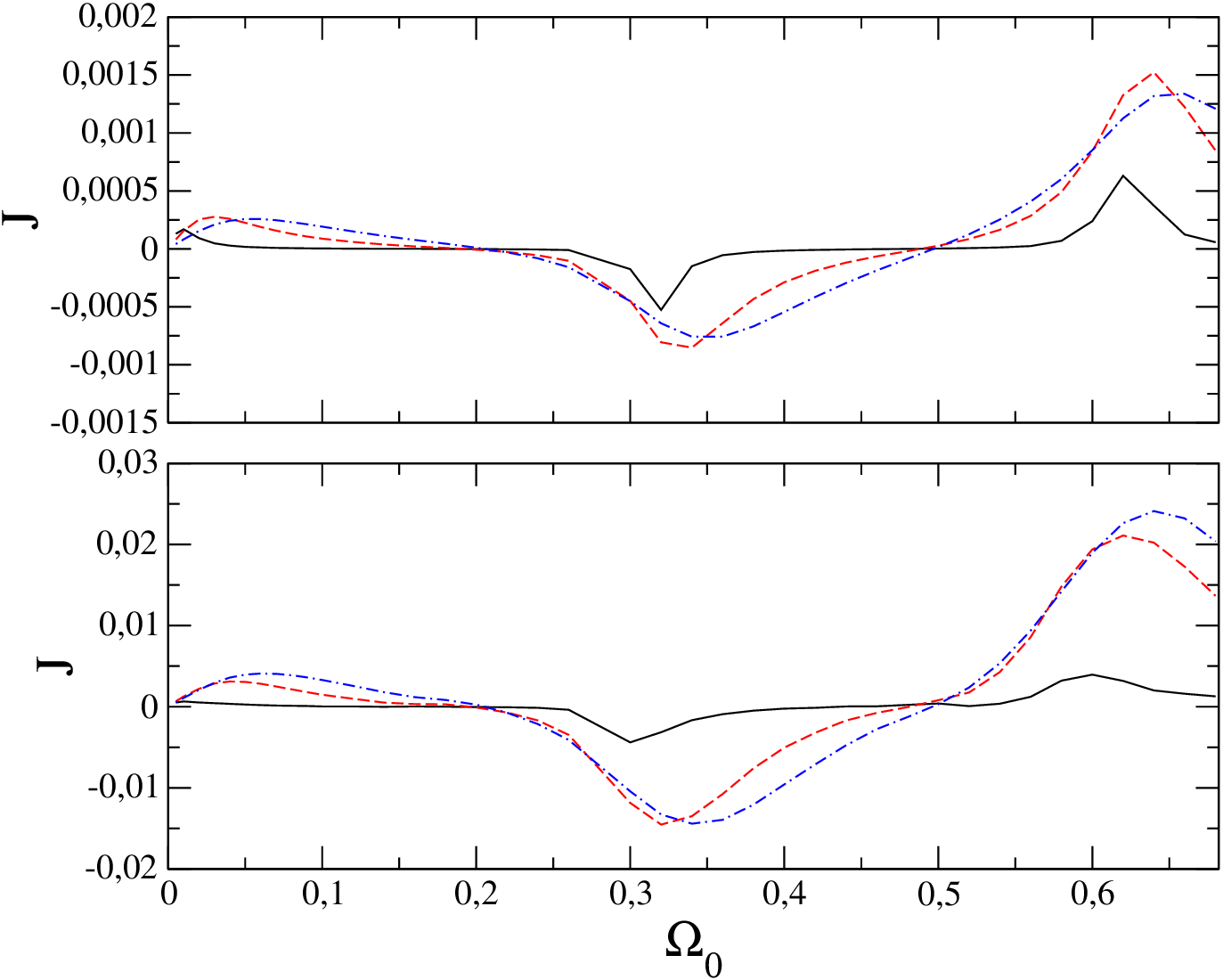}
\caption{(Color online) The direct current $J$ as a function
of the pumping frequency $\Omega_0$ for $\delta=\pi/2$,
chemical potential $\mu=0$,
 and contact parameter 
$ |w_0|^2=0.1,0.5,1$ (black solid, red dashed and blue dot-dashed lines).
Upper and lower panels 
correspond to pumping amplitudes $V=0.2$ and $V=1$, respectively.
Other parameters are as in Fig. \ref{fig4}. 
}
\label{fig8}
\end{figure}
Figure \ref{fig8} shows $J$ as a function of the pumping
frequency $\Omega_0$ for different values of the parameter $|w_0|^2$. The
first feature to note is the structure of minima and maxima corresponding
to resonant frequencies causing interference between nearest and next-nearest 
neighbor energy
levels. The change of sign between resonances
is consistent with the arguments  of Appendix B
and the discussion related to Figs. \ref{fig4} and \ref{fig5}.
Another issue worth mentioning is the linear behavior at very small 
pumping frequencies, as can  be inferred from an expansion of the
low $V$ transmission function
(\ref{tap}) in powers of  $\Omega_0$.

\begin{figure}
\includegraphics[width=8cm,clip]{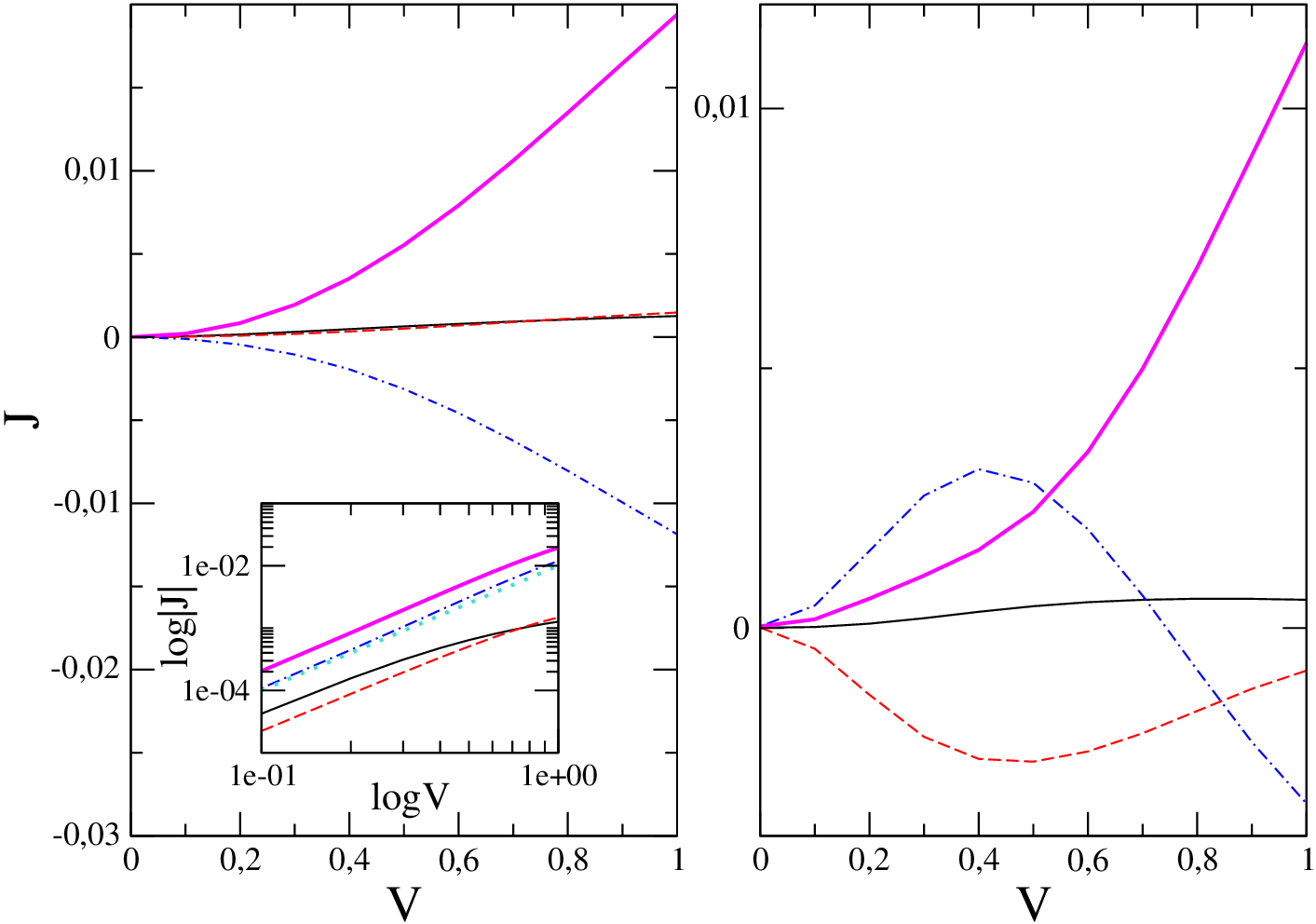}
\caption{(Color online) The direct current $J$ as a function
of the pumping amplitude $V$ for $\delta=\pi/2$ and
$\Omega_0=0.01,0.1,0.3,0.6$ (black thin, red dashed, blue dot-dashed and
magenta thick lines, respectively) and $ |w_0|^2= 0.5$. 
Left and right panels correspond to $\mu=0$ and $\mu=2$, respectively.
The inset in the left panel contains the same data of the main panel
in log-log scale. The light blue dashed line indicates the slope
for $J\propto V^2$.
Other parameters are as in Fig. \ref{fig4}. 
}
\label{fig9}
\end{figure}
To finalize, we show in Fig. \ref{fig9}  $J$ as a function of
the pumping amplitude $V$. 
For chemical potentials bellow the spectral 
edge, the behavior is consistent
with $J \propto V^2$ within a rather wide range of $V$. This can be better observed 
in the log-log plot of the inset.
This law is in agreement with the low $V$ transmission function 
(\ref{tap}).
 Note that this behavior 
is followed at resonance within a wider range of
 $V$  
(see the plots in blue
dot-dashed and thick magenta lines in the left panel).
The right panel shows the situation for chemical potentials close to the
spectral edge. In this case, the behavior for very small $V$ remains in
agreement with the law predicted by (\ref{tap}). For some pumping
frequencies, however, $J$ strongly deviates from this behavior at moderate $V$,
in some cases showing sign reversals. Such dramatic changes have also been
observed in molecular driven ratchets \cite{rat,leh}. Their source
must be found in the higher density of electronic levels at higher energy
and the possibility of interference between more than two 
electronic levels as the pumping amplitude increases.
As in the case of only one harmonically time-dependent potential, the behavior of $T(\omega)$
within the high energy region
has a much higher sensitivity to the particular values of parameters.
  Hence, the prediction of the sign of the current when the chemical
potential tunes this spectral region  turns much more difficult.

\section{Summary and conclusions}
We have presented a general approach based on non-equilibrium Green functions
 to study transport phenomena originated in time-periodic potentials applied on 
quantum systems described by tight-binding Hamiltonians without
many-body interactions. 
The present treatment allows for the exact solution in problems with several
time-dependent potentials provided that the oscillating frequencies are
commensurate (i.e. a multiple of an elementary frequency $\Omega_0$). 
In addition, the present treatment is valid for arbitrary 
amplitudes  and oscillating frequencies of the time-dependent
potentials.

We have employed the general treatment of section II to study two simple 
problems of quantum pumping in a
one-dimensional model for a double barrier structure connected to left and right
reservoirs. We have first considered only one pump acting
on one of the barriers. This case is interesting because an explicit analytical
 expression for the retarded Green function can be found. We have shown that
the existence of a net electronic transport depends exclusively on the
structure of the environment of the pumping center and two simple
conditions must be fulfilled: the geometrical arrangement
must not have spacial inversion symmetry with respect to the pumping center
and it must have resonant levels.

The  second case we have considered corresponds to pumping
potentials oscillating with the same frequency  and a phase-lag between them,
acting at both of the barriers. This kind of operational arrangement of
ac-potentials is just the one used in the experiment of Ref. \cite{swi}.
In particular, we have shown that for weak pumping amplitude, the 
net current behaves like
$J \propto \sin(\delta)$ as a function of the phase lag, in agreement with
the experimental work. Another interesting feature is that for reservoirs
with wide bands (as is often the case in double barrier structures 
in semiconductor junctions), the current can be related to the
difference in the charge density at the pumping points. 

We have stressed the fact that all electrons below the Fermi
energy of the reservoirs contribute to the  current generated by pumping mechanisms.
We have also devoted some effort to understand the conditions that determine
the direction of the net current. We have found that in the two cases
analyzed, this direction coincides with the one predicted by a ``naive'' 
adiabatic picture
in the situations where quantum interference does not play a role.
This condition seems to be more easily achieved in the case of only
one pumping potential. Instead, for the case of two pumping potentials it is
achieved for not too strong pumping amplitudes, when the pumping frequency is smaller than the separation between
consecutive energy levels. For resonant frequencies, sign reversals 
take place due to quantum interference of different electronic levels.
In both examples, there are regions close to the spectral edge where 
the transmission function experiments many changes of signs.
In this region, the details strongly depend on the parameters. 
We have also shown that the current behaves linearly in $\Omega_0$ when the pumping
frequency is small and proportional to $V^2$ for low pumping amplitudes. At
resonance, the latter behavior extends to a range of $V$ beyond the one
expected from arguments based in perturbation theory.

The investigation of the pumping effect in systems with annular topology and
the combination of the time-dependent effects with electron-electron,
electron-phonon interactions and
disorder is left to the future.

\section{Acknowledgments}
The author thanks useful conversations with G. Chiappe, 
L. Mart\'{\i}n-Moreno,
S. Flach,
S. Denysov, as well as the hospitality of Prof Fulde  at the
MPIPKS-Dresden,
where part of this work has been done.
Support from the Alexander von Humboldt Stiftung,
PICT 03-11609 from Argentina, BFM2003-08532-C02-01
from MCEyC of Spain, grant ``Grupo consolidado DGA''
and from the MCEyC of Spain through ``Ramon y Cajal'' program 
 are
acknowledged. LA is staff member of CONICET, Argentina.

\appendix

\section{Evaluation of the retarded Green function for a single
harmonically time-dependent potential.}
We present the solution of the set (\ref{dyret}) for the case of only one
periodic potential. In this case, it is possible to obtain an analytical
expression for $G^R_{1,1}(t,\omega)$ through a recursive procedure. After
some algebra, it is found an expression with the structure (\ref{four2}),
with the following coefficients:
\begin{eqnarray}
{\cal G}_{1,1}(k,\omega) & = & {(\frac{V}{2})}^k g^{eff}(\omega) \prod_{m=1}^k
 g^{(-m)}(\omega - m \Omega_0), \;\;\;\, k>0 \nonumber \\
& = & {(\frac{V}{2})}^{-k} g^{eff}(\omega) \prod_{m=1}^{-k}
 g^{(m)}(\omega + m \Omega_0), \;\;\;\,  k<0 \nonumber \\
{\cal G}_{1,1}(0,\omega) & = & g^{eff}(\omega),
\end{eqnarray}
where
\begin{eqnarray}
g^{eff}(\omega) & = & \frac{1}{\omega- \varepsilon_1- \Sigma^{eff}(\omega)} \nonumber \\
\Sigma^{eff}(\omega) & = & \Sigma^0 (\omega) +
{(\frac{V}{2})}^2 [ g^{(1)}(\omega+ \Omega_0)  \nonumber \\
& & + g^{(-1)}(\omega- \Omega_0) ].
\end{eqnarray}
The ``bare'' self-energy  $\Sigma^0(\omega)= 
|w_L|^2g^R_{L}(\omega) +|w_R|^2g^R_{R}(\omega)$ represents the environment
and is completely defined from the free Green functions of the pieces at the
left ($L$) and at the right ($R$) of the pumping center. 
The function $g^{(m)}(\omega +  m \Omega_0)$ can be expressed as a continued
fraction defined from the recursion relation
\begin{eqnarray}
& & [g^{(m)}(\omega + m \Omega_0)]^{-1}   = 
\omega + m \Omega_0 - \varepsilon_1 - \Sigma^0(\omega + m \Omega_0) \nonumber \\
& & 
-
{(\frac{V}{2})}^2 g^{(m \pm 1)}(\omega + (m \pm 1)  \Omega_0),
\label{rec}
\end{eqnarray}
where $+,- $ corresponds to $m>0$ and $m<0$, respectively. 
In practice, a cut-off is introduced such that
$g^{(\pm K)} (\omega \pm K \Omega_0) = [ \omega \pm K \Omega_0-
\Sigma^0(\omega \pm K \Omega_0)]^{-1}$, being $K$ large enough in 
order to satisfy that $|K\Omega_0|$ is much larger than the
absolute value of the energy for which the bare
Green function $[\omega -\Sigma^0(\omega)]^{-1}$ has non-vanishing spectral weight.  

Using the fact that
\begin{equation}
\mbox{Im}[{\cal G}_{1,1}(0,\omega)]=
 - \mbox{Im}[ \Sigma^{eff}(\omega)]
|g^{eff}(\omega)|^2,
\end{equation}
and the definition of $\Sigma^{eff}(\omega)$, it is found
\begin{equation}
\mbox{Im}[{\cal G}_{1,1}(0,\omega)]=-\sum_k  \mbox{Im}[\Sigma^0(\omega - k \Omega_0) ]
|{\cal G}_{1,1}(k,\omega) |^2.
\label{rel}
\end{equation}

\section{Approximate solution for the 
retarded Green function for  two harmonic
potentials at resonance.}

Let us start by noting that the retarded Green function for a
tight-binding chain with hopping $w$ and length $N$ with open boundary conditions
at both ends is:
\begin{equation}
g^R_{lm}(\omega)=\sum_k \frac{\sin(kl) \sin(km)}{\omega -\epsilon_k + i\eta},
\end{equation}
being $\eta=0^+$, $\epsilon_k= -w \cos(k)$, and $k=n \pi/(N+1)$, 
with $n=1, \ldots, N$. 

We assume that the resonant energies $\epsilon_k$ and phases of the wave function
of the double barrier structure connected to the reservoirs are
approximately those of an open tight-binding chain of the same length
and we propose 
the following ansatz for the retarded Green function
evaluated
at a resonant frequency $\epsilon_k$
\begin{equation}
G^0_{lm}(\epsilon_k) \sim \gamma_k \sin(kl) \sin(km). 
\label{ansatz}
\end{equation}
We have verified that this ansatz reproduces with reasonable accuracy the
phases of the Green function in the present model when the coupling to
the reservoirs is not too weak.

Let us consider weak pumping amplitude and a pumping frequency
such that $\Omega_0 = \epsilon_{k_1}-\epsilon_{k_2}$, being 
$k_1-k_2=j \pi/(N+1) $ being $j$ a
positive integer. The set (\ref{dyretper})
reduces to
\begin{eqnarray}
 G^R_{m,n}(t,\epsilon_{k_1}) & \sim & 
G^0_{m,n}(\epsilon_{k_1})+\frac{V}{2} e^{i (\delta+\Omega_0 t)}
G^0_{m,1}(\epsilon_{k_2})G^0_{1,n}(\epsilon_{k_1}) \nonumber \\
& & + \frac{V}{2} e^{i \Omega_0 t} G^0_{m,N}(\epsilon_{k_2}) 
G^0_{N,n}(\epsilon_{k_1}),
\nonumber \\
G^R_{m,n}(t,\epsilon_{k_2}) & \sim &
G^0_{m,n}(\epsilon_{k_2})+\frac{V}{2}   e^{-i (\delta+\Omega_0 t)}
G^0_{m,1}(\epsilon_{k_1})G^0_{1,n}(\epsilon_{k_2})  \nonumber \\
& & + \frac{V}{2}  e^{-i \Omega_0 t} G^0_{m,N}(\epsilon_{k_1}) 
G^0_{N,n}(\epsilon_{k_2}).
\label{dyretper1}
\end{eqnarray}
The replacement of the ansatz (\ref{ansatz}) in the above equations
reveals that there is a phase equal to $j \pi$ between the
terms proportional to $V e^{\pm i (\delta + \Omega_0 t)}$ in relation to the
ones proportional to $V e^{\pm i  \Omega_0 t} $
in the above expressions for
$ G^R_{m,n}(t,\epsilon_{k_i})$. This indicates that the effective
phase-lag between the two potentials is $\delta+ j \pi$.

\end{document}